\documentclass[aps,prl,amsmath,amssymb,amsfonts,floatfix,reprint]{revtex4-1}
\usepackage{graphicx}
\usepackage[usenames, dvipsnames]{color}
\pdfoutput=1

\begin{document}

\title{Magnon scattering in the transport coefficients of CoFe alloys}

\author{S. Srichandan$^{1}$}
\author{S. Wimmer$^{2}$}
\author{M. Kronseder$^{1}$}
\author{H. Ebert$^{2}$}
\author{C.~H. Back$^{1}$}
\author{C. Strunk$^{1}$}

\affiliation{$^{1}$Institute of Experimental and Applied Physics, University of Regensburg, D-93040, Germany}
\affiliation{$^{2}$Department of Chemistry, Physical Chemistry, Ludwig-Maximilians-Universit\"at, Munich, D-81377, Germany}
\date{\today}

\begin{abstract}
	Resistivity $\rho$, thermopower ${\cal S}$, and thermal conductivity $\kappa$ were measured simultaneously on a set of CoFe alloy films. Variation of the Co-content $x_\mathrm{Co}$ allows for a systematic tuning of the Fermi level through the band structure, and the study of the interplay between electronic and magnetic contributions to the transport coefficients. While band structure and magnon effects in $\rho$ and $\kappa$ are rather weak, they turn out to be very significant in ${\cal S}$. The evolution of Mott and magnon drag contributions to ${\cal S}$ is traced  between the two limiting cases of pure Fe and pure Co.  In addition, we find an interesting sign change of the magnon drag.
\end{abstract}

\maketitle

Spintronics \cite{Zutic,Ralph} and more recently spin-caloritronics \cite{Bauer, Boona} have sparked interest in the fundamental transport properties of ferromagnetic thin films since devices engineered from ultra thin ferromagnetic layer stacks have potential for technological applications. While the measurement and interpretation of electrical transport parameters is rather straight forward even for thin ferromagnetic films \cite{Pratt}, the measurements and interpretation of their thermal, thermo-electric and magneto-thermo-electric counterparts is much more difficult. However, the optimization of spintronic and spin-caloritronic devices depends on the accurate knowledge of the various thermal transport parameters as well as the parameters governing the relaxation mechanisms for electrons, phonons and magnons in thin film ferromagnetic materials.  Similarly, the exploitation of magnon transport in temperature gradients for transmission and processing of information \cite{Nikitov,Cornelissen} depends on the understanding and quantitative knowledge of their thermoelectric and thermomagnetic properties.


So far only a few experiments have addressed the interplay of the  magneto-thermo-electric transport parameters
 using the modern tool-box of nanotechnology \cite{Denlinger,Lope,Ou,Zink_solid,Cooke,Avery,Schmid,Zink}; these were mainly focused on the prototypical ferromagnet permalloy while systematic investigations as a function of alloy composition are still lacking.
On the theory side, significant progress  has been made in the description of spin dependent transport phenomena. The use of \textit{ab initio} theory in combination with a realistic description of alloys \cite{Ebert,Kovacik,Obstbaum,Meyer} allows now for a fresh look on the transport properties of ferromagnetic alloys. Of particular interest is the prediction of Flebus {\it et al.} \cite{Flebus}, who pointed out that besides the usual diffusion term in the thermoelectric power (TEP) two contributions compete in the {\it magnon drag}: one of hydrodynamic origin that drives majority carriers towards the cold side of the sample, and a second one in the opposite direction. The second contribution arises from the accumulation of spin-Berry phase in a time-dependent magnetization texture \cite{Tserkovnyak}, caused here by the thermally excited spin waves.


Experimental evidence for magnon drag effects in the TEP has been reported for elemental Fe \cite{Blatt}  and Cr \cite{behnia} bulk samples.  Only very recently the topic was taken up again by Watzman et al.  \cite{Watzman}, who attributed an important contribution to the TEP and the Nernst coefficient of elemental Fe and Co to magnon scattering. Interestingly, the sign of the presumed magnon contribution to the TEP is opposite for both metals. Hence the  natural questions arise, what is the reason for this sign change and what is the evolution of the TEP in CoFe alloys between the two elements. With varying composition not only the electron density, but also the phonon and magnon dispersion relations change. This affects all sources of scattering processes for the electrons and thus the temperature dependence of the transport coefficients. So far only the electric and spin transport in CoFe alloys were recently carefully studied, and the spinwave damping parameters $\alpha(x_\mathrm{Co})$ measured \cite{schoen1, schoen2}. 


In this Letter we investigate a series of CoFe alloy films on SiN$_{x\,}$-based suspended microcalorimeters. Simultaneous measurements of several transport coefficients, i.e., the resistivity $\rho(T)$, the TEP ${\cal S}(T)$ and the thermal conductivity $\kappa(T)$ are performed in a wide temperature range of 25-300 K on the very same films. In this way, we directly probe the variation of the spin-polarized band structure and the relevant scattering mechanisms with the Co-content, and the evolution of magnon scattering in different observables. We find evidence for magnon scattering effects most clearly in the TEP.  The magnon drag contribution ${\cal S}_\mathrm{mag}(T,x_\mathrm{Co})\propto T^{3/2}$ systematically decreases with $x_\mathrm{Co}$, and changes sign near $x_\mathrm{Co} \simeq 0.6$.

\begin{figure}
		\includegraphics[width=8cm]{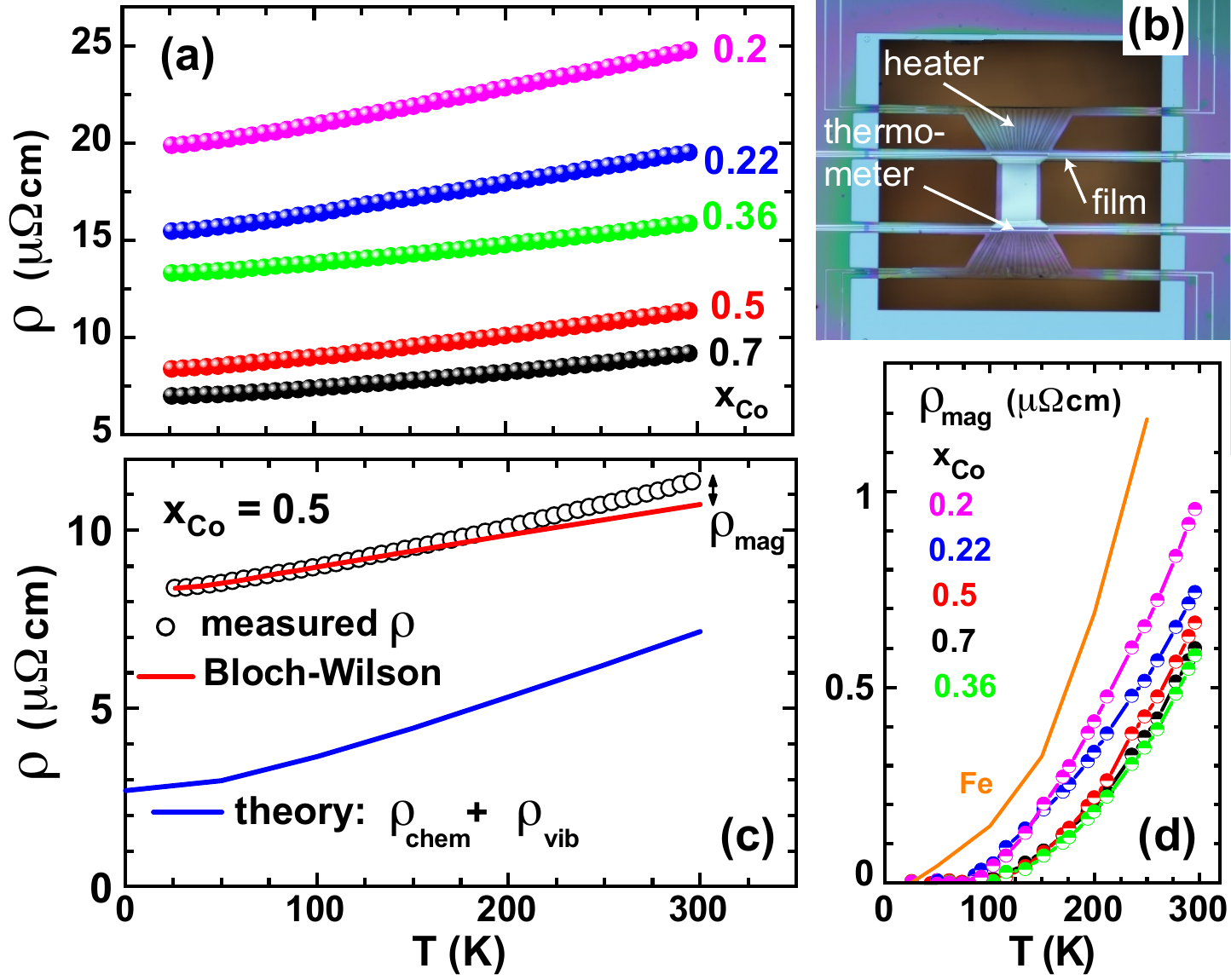}
%
		\caption{(a)~Resistivity $\rho(T)$ for several values of Co content $x_\mathrm{Co}$.
(b) Optical image of a typical device.		(c)~The phonon contribution to the resistivity for $x_\mathrm{Co}=0.5$ (open circles) fitted to a Bloch-Wilson function for $T\leq100$\,K (red line). Double headed arrow: estimated magnon contribution to $\rho(T)$. Blue line: calculated resistivity taking into account chemical disorder and lattice vibrations for for $x_\mathrm{Co}=0.5$ (see text). (d)~Estimated $\rho_\mathrm{mag}$ for all samples (dots) vs.~temperature with $\rho_\mathrm{mag}$ for bulk Fe from Ref.~\onlinecite{Raquet} (orange line). }
\label{Fig1}		
%
\end{figure}
To fabricate the samples, 60-80 nm thick CoFe films are deposited as rectangles (116\,$\mu$m $\times$ 60\,$\mu$m) by molecular beam epitaxy in an ultra high vacuum chamber on  500\,nm thick SiN$_x$ membranes [light blue in Fig.~\ref{Fig1}(b)] with an area of 500\,$\mu$m $\times$ 500\,$\mu$m. The film is examined using atomic force microscopy for the determination of the surface roughness, by X-ray photo-electron spectroscopy for stoichiometry determination, and by X-ray diffractometry. The crystal structure of alloys with $x_\mathrm{Co}$ = 0.2 and 0.22 turn out to be predominantly {\it bcc} while the films with $x_\mathrm{Co}$ = 0.36, 0.5 and 0.7 display traces of {\it fcc}-precipitations, similar to the findings in Ref.~\onlinecite{schoen1}. 

Next contact leads and thermometers are patterned using e-beam lithography (EBL) and deposition of $50$\,nm of Al. The thermometers are 100\,$\mu$m long and 1.5\,$\mu$m wide wires. The contact leads are also 1.5\,$\mu$m wide. In a second EBL step two symmetrically placed meander heater structures are patterned in a $40$~nm thick Au$_\mathrm{60}$Pd$_\mathrm{40}$ film. Finally, the parts of the membranes that do not support the metal structures  [black area in Fig.~\ref{Fig1}(b)] are  reactively etched using a CHF$_3$ / O$_2$ plasma for 10 minutes, leaving a freely suspended SiN$_x$-bridge. For more details see the Supplementary Material \cite{supplement}). 

The measurements were performed in a helium flow cryostat in vacuum. Radiation losses are minimized by virtue of a radiation shield at the sample temperature. All resistances were measured in a four terminal configuration. The TEP ${\cal S}(T)$ and the thermal conductance $K(T)$ were determined simultaneously by measuring the temperature difference $\Delta T$ between the ends of the bridge vs.~heater current such that  $\Delta T/T<0.01$. The corresponding thermovoltage $V_\mathrm{th}$ is measured using a nanovoltmeter and the TEP is extracted from the slope  of $V_\mathrm{th}(\Delta T)$. 
The total thermal conductance $K = P_\mathrm{H}/\Delta T$ includes the thermal conductances $K_\mathrm{B}$ and $K_\mathrm{L}$ of the bridge and the lead  sections, respectively.  $P_\mathrm{H}$ is the heater power. In the absence of radiation or convection losses, the 1D heat diffusion equation can be solved to find $K_\mathrm{B}$ and $K_\mathrm{L}$ independently \cite{Kim, Zink}.  $K_\mathrm{B}$ contains both $K_\mathrm{CoFe}$ and  $K_{\mathrm{SiN}_{x}}$. To determine $K_\mathrm{SiN}$, we have prepared 4 devices with bare SiN$_{x}$. From the thermal conductance $K_\mathrm{CoFe} = K_\mathrm{B} - K_{\mathrm{SiN}_x}$ we calculate the thermal conductivity $\kappa_\mathrm{CoFe} = \kappa$ using the known dimensions of the film for all the samples with different compositions. The uncertainty of $\kappa$ resulting from the variance of $K_\mathrm{SiN}$ between the different SiN$_{x}$ membranes is $\simeq7.5$\,W/(K\,m).

\begin{figure*}
		\includegraphics[width=12cm]{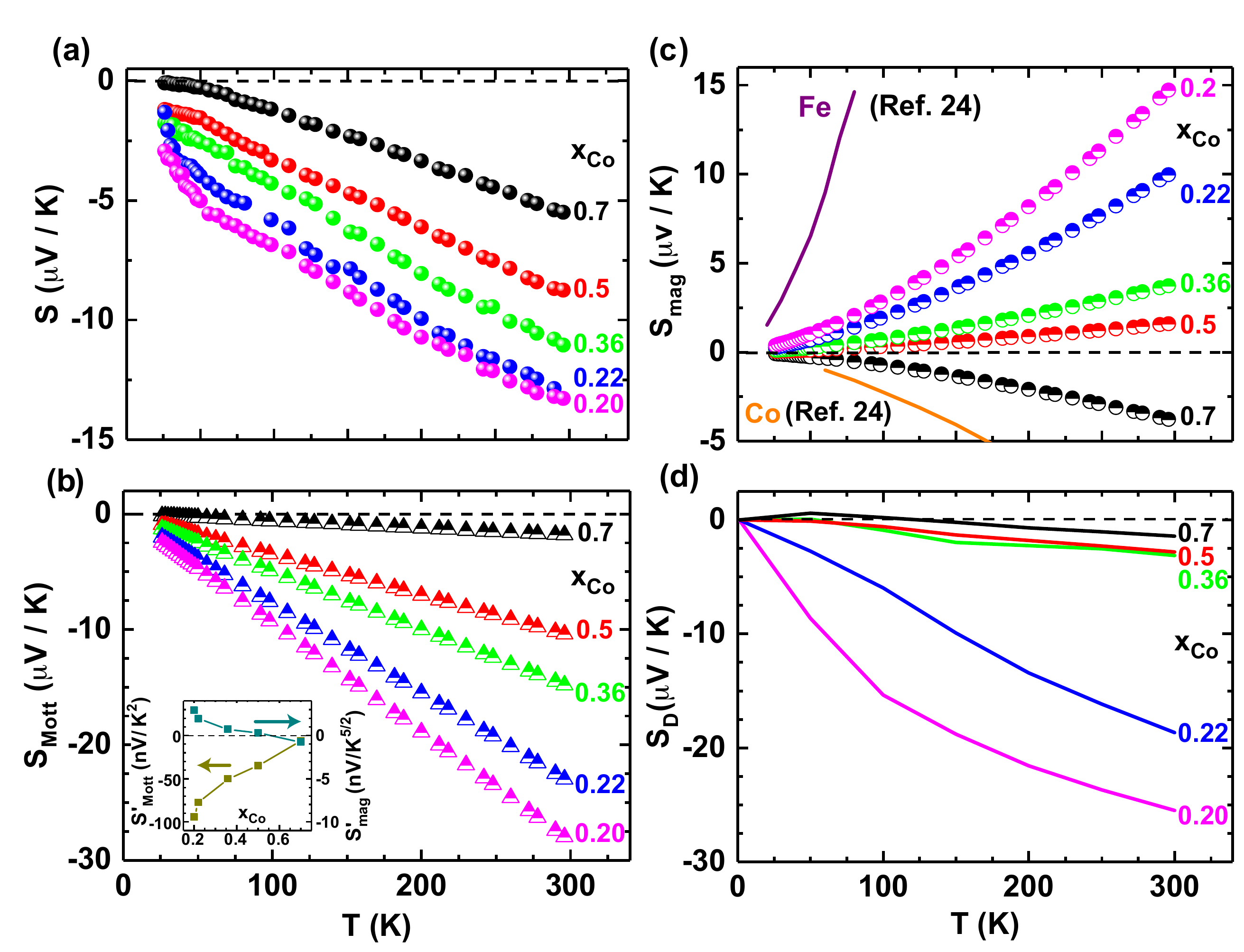}
%
		\caption{(a)~Measured thermopower vs.~temperature for all Co$_x$Fe$_{1-x}$ samples from $26$~K-$296$~K, labeled on the right by at$\%$  Co. (b) Mott-like contribution ${\cal S}_\mathrm{Mott}(T)$ (half filled triangles) as a function of temperature. Inset: ${\cal S}_\mathrm{D}$ and ${\cal S}_\mathrm{theo}$ vs.~at$\%$ Co at 296 K. c) Magnon drag ${\cal S}_\mathrm{mag}(T)\propto T^{3/2}$ contribution (half filled circles), literature values for bulk Co and Fe \cite{Watzman} are shown as lines. (d) Calculated thermopower taking into account chemical  and vibrational disorder.}
		\label{Fig2}
%
\end{figure*}

In Fig.~\ref{Fig1}(a) the resistivity $\rho(T)$ of all five samples is plotted as a function of temperature. 
The resistivity is highest for $x_\mathrm{Co} = 0.20$ and  decreases monotonically with addition of Co.
This  decrease is mainly a consequence of the increase in electron number. In addition, at $x_\mathrm{Co} \simeq 0.2$,  a $d$-like band crosses the Fermi surface, resulting in a  maximal $\rho(x_\mathrm{Co})$ (for more details and a comparison with earlier experiments see the Supplementary Material \cite{supplement}).

Next we evaluate the magnon contribution $\rho_\mathrm{mag}$ to $\Delta\rho(T)$.  According to the analysis of Ref.~\cite{Venkat,Raquet}  $\rho_\mathrm{mag}$ becomes sizable only above $T \simeq 100$\,K.  Hence, we first determine the phonon contribution by fitting the measured $\rho(T)$  to a Bloch-Wilson (BW) function \cite{supplement} from 26\,K up to 100\,K. An example is shown in Fig.~\ref{Fig1}(c) for $x_\mathrm{Co} = 0.5$.  Extrapolating to 300\,K we can evaluate the magnon contribution $\rho_\mathrm{mag}(T)$  by subtracting the BW-fit from  the  measured $\rho(T)$. The results are plotted in Fig.~\ref{Fig1}(d):  $\rho_\mathrm{mag}(T)$  gradually decreases with increasing $x_\mathrm{Co}$ (with  $x_{Co} = 0.36$ being an outlier). The magnon contribution is at most $6.5\%$ of $\rho$ at room temperature for $x_\mathrm{Co} = 0.20$, corresponding to about $1/5$ of the phonon contribution. The magnitude and temperature dependence of $\rho_\mathrm{mag}(T)$ are quite comparable to that of elemental Fe \cite{Raquet} [orange line in Fig.~\ref{Fig1}(d)].
The blue line in Fig.~\ref{Fig1}(c) shows a first-principles calculation of $\rho(T)$ for $x_\mathrm{Co}= 0.5$ within the Kubo formalism accounting for chemical disorder via the CPA alloy theory and for thermal lattice vibrations via the Alloy Analogy Model~\cite{Ebert}. The calculation underestimates the absolute values and overestimates the slope of $\rho(T)$ both by a factor of $\simeq 2$ as it does not include the considerable structural disorder.

Next we present the results for the thermopower in Fig.~\ref{Fig2}(a), which constitutes our main result. At high temperatures ${\cal S}(T)$ is negative and varies roughly linearly with temperature. Note that the approximately linear parts at $T>100\,$K do not extrapolate to  ${\cal S} = 0$ at $T = 0$, as opposed to the expectation from the Mott-law. At low temperatures  ${\cal S}(T)$ is not linear. This implies that ${\cal S}(T)$ cannot be described by a Mott-like dependence alone, but additional non-linear contributions have to be present. Moreover, the curvature clearly changes sign: it is  positive for lower Co content, i.e., $x_\mathrm{Co}$ = 0.2 and 0.22, but negative for $x_\mathrm{Co} = 0.7$ and $0.5$. At the lowest temperatures ${\cal S}(T,x_\mathrm{Co} = 0.7)$ becomes slightly positive. 

By fitting the high temperature part of ${\cal S}(T)$ to a Mott-like term linear in $T$, and a second term proportional to $T^{3/2}$, we can decompose the TEP according to
\begin{equation}\label{TEP}
{\cal S}(T) = {\cal S}^{'}_\mathrm{Mott}\cdot T+{\cal S}^{'}_\mathrm{mag} \cdot T^{3/2} + {\cal S}_\mathrm{res}(T)\;.
\end{equation}
The coefficients ${\cal S}^{'}_\mathrm{Mott}(x_\mathrm{Co})$ and ${\cal S}^{'}_\mathrm{mag}(x_\mathrm{Co})$ describe the dependencies of the Mott-like part ${\cal S}_\mathrm{Mott}(T)$ 
and magnon drag contribution ${\cal S}_\mathrm{mag}(T)$ 
on $x_\mathrm{Co}$. We have verified that these coefficients are robust against a change of the fit interval within 100-300\,K.
 Below 100\,K a much smaller residual contribution ${\cal S}_\mathrm{res}(T)\lesssim1\mu$V/K remains  (see Supplementary Material \cite{supplement}).

 Figure~\ref{Fig2}(b) shows the Mott-like contribution that is proportional to $T$. The absolute values $|{\cal S}_\mathrm{Mott}(T)|$ decrease with increasing Co-content, i.e., with increasing electron density, which is consistent with the corresponding trend seen in $\rho(T)$.  The values of ${\cal S}^\prime_\mathrm{D}$ contain a small contribution  ${\cal S}^\prime_\mathrm{D, Al} = 3.7\,$nV/K$^2$ from the diffusion thermopower of the Al leads \cite{Huebener}.

 On the other hand,  we find a substantial nonlinear contribution $S_\mathrm{mag}(T)$ that increases proportional to the magnon number and is as large as 13.5 $\mu$V/K at 296 K for the film with $x_\mathrm{Co} = 0.2$ [Fig.~\ref{Fig2}(c)].  The sign of the coefficient ${\cal S}^{'}_\mathrm{mag}$ is positive for $x_\mathrm{Co} \lesssim 0.5$ and negative for $x_\mathrm{Co}$ = 0.7 (inset). This is reflected in the sign change of ${\cal S}_\mathrm{mag}(T)$ from positive for the Fe-rich to negative for the Co-rich alloys, which agrees with ${\cal S}_\mathrm{mag}$ for the case of elemental Fe and Co \cite{Watzman} at these temperatures. The inset in Fig.~\ref{Fig2}(b) shows the evolution of the coefficients ${\cal S}^{'}_\mathrm{Mott}$ and ${\cal S}^{'}_\mathrm{mag}$ with $x_\mathrm{Co}$.

 In ferromagnets, the magnon drag contribution to the TEP has a $T^{3/2}$ dependence  at low $T$, provided that $T>\Delta_\mathrm{mag}/k_\mathrm{B}$ ($\Delta_\mathrm{mag}$ being the gap in the magnon dispersion relation),  which reflects the variation of magnon density and specific heat with $T$. The magnon drag peak normally occurs at a temperature roughly one fifth to one half of the Curie temperature $T_\mathrm{C}$ of the material  \cite{Blatt}. Due to the high $T_\mathrm{C}$ of the studied CoFe alloys the maximal the magnon drag for our films is expected above the  temperature range investigated here. 
 
 In Fig.~\ref{Fig2}(d) we show the  diffusion contribution, ${\cal S}_\mathrm{D}$, to the TEP, as obtained from first-principles calculations. For the highest and lowest Co-concentration the calculation can reproduce the size and systematics of the experimental data, but for intermediate concentrations it significantly underestimates both the measured TEP in Fig.~\ref{Fig2}(a) and the linear contribution to the TEP in Fig.~\ref{Fig2}(b). In theory the curvature arises from the rapid variation of the energy-dependent conductivity when the $d$-bands touch the Fermi energy around $x_\mathrm{Co}\simeq0.2$  (see Supplementary Material for details \cite{supplement}). At high temperatures, this requires to go beyond the term linear in $T$ in the Sommerfeld expansion.  Taking into account also spin disorder further reduces ${\cal S}_\mathrm{D}$. Given the significant curvature of the measured thermopower below 100\,K, our experimental results cannot be explained by the diffusion contribution alone.  
  The computed suppression  of ${\cal S}_\mathrm{D}(x_\mathrm{Co} \lesssim 0.5)$ can, in part, be reverted by the presence of {\it fcc}-precipitations with intrinsically larger absolute values of ${\cal S}_\mathrm{D}$ and an opposite curvature \cite{supplement,footnote}. The relevance of such precipitations is also corroborated by the behavior or the thermal conductivity (see below).
 

Most interesting is the sign change observed for ${\cal S}_\mathrm{mag}$ when $x_\mathrm{Co}$ is tuned from the Fe to the Co-rich side. As already mentioned, recent theoretical work has calculated the spin-motive forces in presence of a magnetization texture \cite{Flebus}:
 (i) a Berry-phase contribution that drives the majority spins towards the hot end and is controlled by the adiabatic damping parameter $\beta$, and (ii) a hydrodynamic contribution that drives the majority spins towards the cold end and is controlled by the Gilbert damping $\alpha$. A finite difference between majority and minority spin-motive force results in an electromotive force proportional to the magnon number (i.e., $\propto T^{3/2}$). The magnetic texture induced by thermally excited magnon generates a magnon drag contribution to the TEP. The Gilbert damping $\alpha(x_\mathrm{Co})$  has been determined from ferromagnetic resonance experiments \cite{schoen2}.  
So far the analysis of our data using this strongly simplified model results in unphysically high values of $\beta$. On the other hand, the clear systematics that we observe calls for a more quantitative theoretical treatment of magnon drag.

\begin{figure}
\centering
\includegraphics[width=8.5cm]{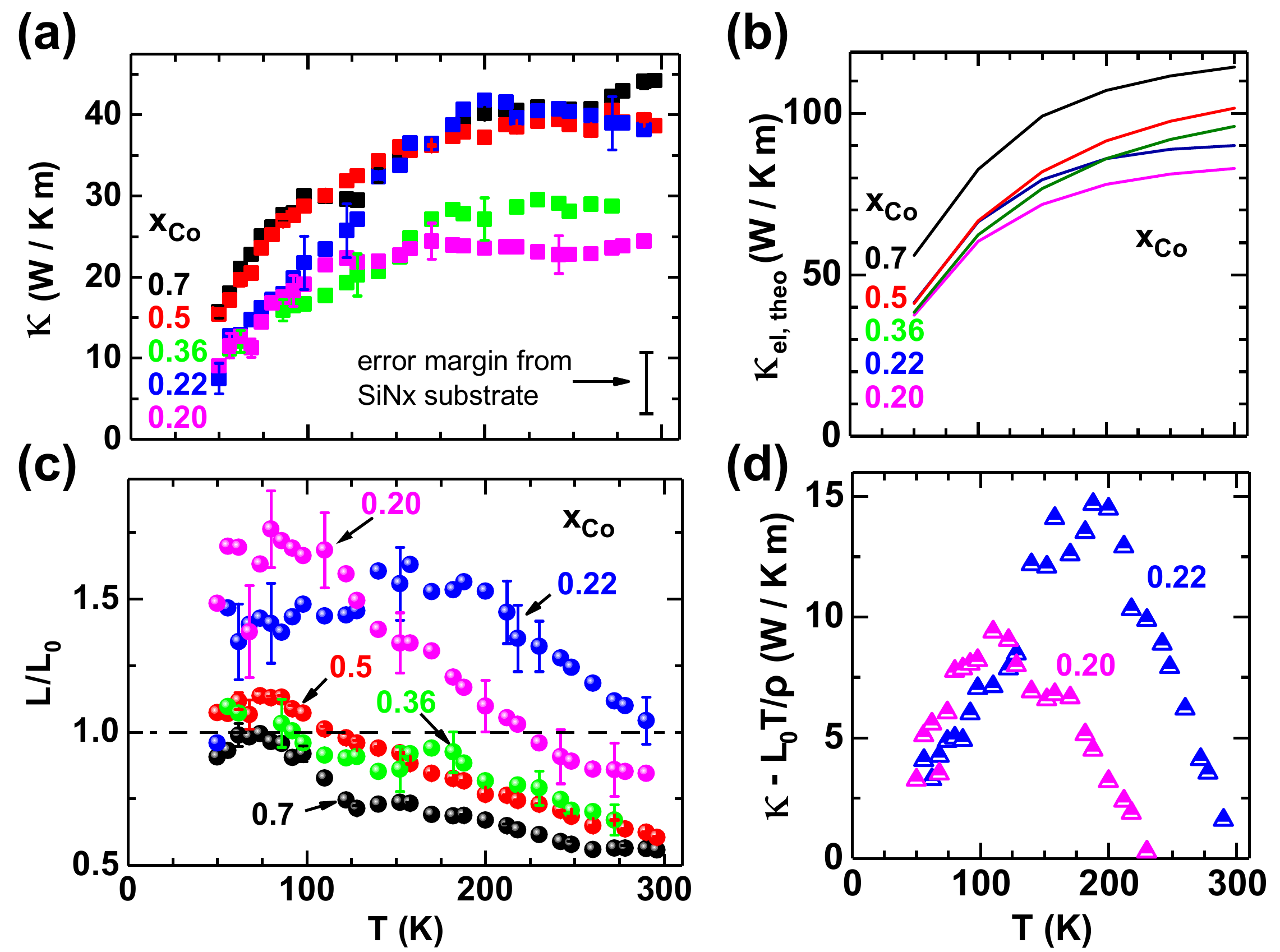}
\caption{(a)~Measured thermal conductivity of the Co$_x$Fe$_{1-x}$-films. (b)~Calculated electronic contribution to $\kappa_\mathrm{el}$ including elastic scattering on lattice vibrations. (c)~Extracted Lorenz number $L$ as a function of bath temperature. The horizontal line indicates the Sommerfeld value  $L_0= \pi^2/3\cdot (k_\mathrm{B}/e)^2$. (d)~Positive deviations from the Wiedemann-Franz-law using the measured resistivity of the films with low Co-content.} 
\label{Fig3}
\end{figure}

Finally, we investigate the thermal conductivity  $\kappa$ in the films. As demonstrated in Fig.~\ref{Fig3}(a) $\kappa(T)$ increases with temperature and then saturates at high temperatures for all films. The individual curves are subjected to a $\pm10\%$ random shift from the slightly varying background contribution of the different SiN$_x$ membranes (see Supplementary Material \cite{supplement}). The corresponding calculation of the electronic contribution $\kappa_\text{el}(T)$ including temperature-dependent vibrational disorder in Fig.~\ref{Fig3}(b) overall reproduces the systematics and the proportions for samples of different Co-contents, with the exception that the monotonic increase of $\kappa$ with $x_{\rm Co}$ observed in the calculated data is violated for $x_{\rm Co} = 0.36$ at high $T$ in our experiment. The absolute values of $\kappa$ are overestimated by the very same factor of $\simeq 2$, by which the theory underestimates the electric resistivity in Fig.~\ref{Fig1}(c). 

The Lorenz number $L(T) = \kappa(T)\rho(T)/T$ evaluated from the measured set of $\rho$ and $\kappa$ is shown in Fig.~\ref{Fig3}(c). We observe a significant violation of Wiedemann-Franz law (WFL, indicated by the horizontal line). Enhancement of $L$ above  $L_0 $ is found for $x_\mathrm{Co}$ = 0.22, while $L$ is smaller than $L_{0}$ for $x_\mathrm{Co}$ = 0.7 at all temperatures. For intermediate $x_\mathrm{Co}$, $L>L_0$ at low $T$ and vice versa at higher $T$. The positive deviation from WFL, i.e., $L > L_{0}$ is naturally explained by the contribution $\kappa_\mathrm{ph}$ from phonons to the thermal conductivity.   In the investigated $T$-regime the magnon contribution to $\kappa$ is usually small compared to the phonon contribution \cite{Edmonds}.  Only in films without {\it fcc} precipitations ($x_\mathrm{Co}$ = 0.2 and 0.22)  one expects  $\kappa_\mathrm{ph}$ to be significant, because such precipitations drastically shorten the phonon mean free path. Hence we estimate $\kappa_\mathrm{ph}\simeq\kappa-T L_{0}/\rho$ (see Fig.~\ref{Fig3}(d); Supplementary Material \cite{supplement}); it shows clear maxima around 100~K and 200~K, respectively, which  resemble the well known Umklapp peak.  They are shifted towards higher temperatures with respect to the phononic Umklapp peak for pure Fe or Co. 


The  observed negative deviations from WFL can be explained by the very short phonon mean free path in films with $x_\mathrm{Co}\geq 0.36$. Besides suppressing $\kappa_\mathrm{ph}$, the electronic contribution $\kappa_\mathrm{el}$ is known to be enhanced in the presence of inelastic ('vertical')  scattering of electron with phonons \cite{Avery, Tritt}, while these scattering events are not effective in the resistivity. In addition, it is known that  $L<L_0$ for pure Co in this temperature range \cite{Laubitz}, and is  thus in agreement with the behavior of $L(T)$ in Co-rich samples. 

To summarize, simultaneous measurement of the electric, thermoelectric, and thermal transport coefficients performed on alloyed CoFe films have enabled us to understand contribution from electrons, phonons and magnons qualitatively and in part even quantitatively.  In particular, the thermopower displays a  competition of diffusion and magnon drag contributions. The magnon drag contribution varies smoothly between the  limits of pure Fe and Co, changing sign close to the center of the concentration range. For the thermal conductivity a pronounced violation of the Wiedemann-Franz law is observed in structurally homogeneous samples with low Co content. A quantitative understanding of the observed systematic evolution of  magnon drag calls for more elaborate theory.
 
 \begin{acknowledgments}
The authors thank T.N.G. Meier for the XPS analysis, M. Zimmermann for AFM measurements, M. Vogel for COMSOL simulations,  C. S\"{u}rgers for the x-ray characterization of the CoFe films, Y. Tserkovnyak and R.\,A. Duine for helpful comments on their theory, and gratefully acknowledge financial support by the Deutsche Forschungsgemeinschaft (DFG) within the priority programme SpinCaT (SPP 1538) and the Bundesministerium f\"ur Bildung und Forschung (BMBF).
 \end{acknowledgments}
\appendix


\begin{thebibliography}{99}

\bibitem[]{Zutic} I. Zuti\'c, J. Fabian and S. D. Sharma \textit{Rev. Mod. Phys.} \textbf{76}, 323 (2004).

\bibitem[]{Ralph} D. C. Ralph and M. D. Stiles \textit{J. Magn. Magn. Mater.} \textbf{320}, 1190-1216 (2008).

\bibitem[]{Bauer} G.E.W. Bauer, A. H. MacDonald and S. Maekawa \textit{Sol. Stat. Commun.} \textbf{150}, 459 (2010).

\bibitem[]{Boona} S. R. Boona,  Vlaminck, R. C. Myers and J. P. Heremans \textit{Energy Environ. Sci.} \textbf{7}, 885 (2014).

\bibitem[]{Pratt}  S. Maekawa and T. Shinjo, 
\textit{Adv. Cond. Mat. Sci.} {\bf 3} (CRC press LLC 2002). 


\bibitem[]{Nikitov} S. A. Nikitov, {\it et al.}, 
\textit{Physics-Uspekhi} {\bf 58} , 10 (2015).


\bibitem{Cornelissen} L.J. Cornelisen, J. Liu, R.A. Duine, J.B. Youssef, and B.J. van Wees, \textit{Nat. Phys.} \textbf{11}, 1022 (2015).



\bibitem[]{Denlinger} D. W. Denlinger, E. N. Abarra, K. Allen, P. W. Rooney, M. T. Messer, S. K.Watson and F. Hellman,
\textit{ Rev. Sci. Instrum.} \textbf{65}, 946 (1994).

\bibitem[]{Lope} A. Lopeandia, L. Cerdo, M. Clavaguera-Mora, L. R. Arana, K. Jensen, F. Munoz, and J. Rodriguez-Viejo, \textit{Rev. Sci. Instrum.} \textbf{76}, 065104 (2005).

\bibitem[]{Ou} M.N. Ou, T.J. Yang, S.R. Harutyunyan, Y.Y. Chen, C.D. Chen, S.J. Lai, \textit{Appl. Phys. Lett.} \textbf{92}, 063101 (2008).

\bibitem[]{Zink_solid} B.L. Zink, A.D. Avery, Rubina Sultan, D. Bassett, M.R. Pufall, \textit{Solid State Commun.} \textbf{150}, 514 (2010).

\bibitem[]{Cooke} D.W. Cooke, F. Hellman, J.R. Groves, B.M. Clemens, S. Moyerman, \textit{Rev. Sci. Instrum.} \textbf{82}, 023908 (2011).

\bibitem[]{Avery} A. D. Avery, M. R. Pufall, and B. L. Zink, \textit{Phys. Rev. Lett.} \textbf{109}, 196602 (2012).

\bibitem[]{Schmid} M. Schmid, S. Srichandan, D. Meier, T. Kuschel, J.-M. Schmalhorst, M. Vogel, G. Reiss, C. Strunk, and C. H. Back, \textit{Phys. Rev. Lett.} \textbf{111} 187201 (2013).

\bibitem[]{Zink} R. Sultan, A. D. Avery, G. Stiehl, and B. L. Zink \textit{J. Appl. Phys.} \textbf{105}, 043501 (2009).

\bibitem[]{Ebert} H. Ebert, S. Mankovsky, K. Chadova, S. Polesya, J. Min\'ar, and D. K\"{o}dderitzsch, \textit{Phys. Rev. B} \textbf{91}, 165132 (2015).

\bibitem[]{Kovacik} R. Kov\'{a}\v{c}ik, P. Mavropoulos, and S. Bl\"{u}gel, \textit{Phys. Rev. B} \textbf{91}, 014421 (2015).

\bibitem[]{Obstbaum} M. Obstbaum, M. Decker, A. K. Greitner, M. Haertinger, T. N. G. Meier, M. Kronseder, K. Chadova, S. Wimmer, D. K\"{o}dderitzsch, H. Ebert, and C. H. Back, \textit{Phys. Rev. Lett.} \textbf{117}, 167204 (2016).

\bibitem[]{Meyer} S. Meyer, Y.-T. Chen, S. Wimmer, M. Althammer, T. Wimmer, R. Schlitz, S. Gepr\"ags, H. Huebl, D. K\"{o}dderitzsch, H. Ebert, G. E. W. Bauer, R. Gross, and S. T. B. Goennenwein, \textit{Nat. Mater.} \textbf{16}, 977 (2017).

\bibitem{Flebus} B. Flebus, R. A. Duine, Y. Tserkovnyak, \textit{Europhys. Lett.}  \textbf{115}, 57004 (2016).

\bibitem{Tserkovnyak}
Y.~Tserkovnyak, C.~H.~Wong, \textit{Phys. Rev. B} \textbf{79}, 014402 (2009).

\bibitem[]{Blatt} F. J. Blatt, D. J. Flood, V. Rowe, P. A. Schroeder and J. E. Cox \textit{Phys. Rev. Lett.} \textbf{18} 395 (1967).

\bibitem[]{behnia} A.~L.~Trego, A.~R.~Mackintosh, Phys.~Rev.~{\bf 166}, 495 (1968).

\bibitem[]{Watzman} S. J. Watzman, R. A. Duine, Y. Tserkovnyak, S. R. Boona, H. Jin, A. Prakash, Y. Zheng and J. P. Heremans \textit{Phys. Rev. B} \textbf{94}, 144407 (2016).



\bibitem{schoen1} M. A. W.  Schoen, J. Lucassen, H. T. Nembach,  T. J. Silva, B. Koopmans, C. H. Back, J. M. Shaw, \textit{Phys. Rev. B} \textbf{95}, 134410 (2017).

\bibitem{schoen2} M. A. W.  Schoen, J. Lucassen, H. T. Nembach, B. Koopmans, T. J. Silva, C. H. Back, J. M. Shaw, \textit{Phys. Rev. B} \textbf{95}, 134411 (2017).

\bibitem{supplement} Supplementary Material at ... 

\bibitem[]{Kim} P. Kim, L. Shi, A. Majumdar and P. L. McEuen \textit{Phys. Rev. Lett.} \textbf{87} 215502 (2001).






\bibitem[]{Raquet} B. Raquet, M. Viret, E. Sondergard, O. Cespedes and R. Mamy \textit{Phys. Rev. B} \textbf{66}, 024433 (2002).


\bibitem[]{Venkat} M. V. Kamalakar, A. K. Raychaudhuri, X. Wei, J. Teng and P. D. Prewett \textit {Appl. Phys. Lett.} \textbf{95}, 013112 (2009).



\bibitem{Huebener} R.~P.~Huebener,
Phys.~Rev.~{\bf 171}, 634 (1968).
%
%
%
%
%



%




\bibitem{footnote}
In the fcc structure and for intermediate Co concentration the Fermi level is situated more closely above the step, while $\sigma(E)$ increases more strongly, when compared to the bcc-structure. At elevated $T$ the step contributes to $\cal{S}(T)$, resulting in an opposite curvature.

\bibitem{Edmonds}
D.~T.~Edmonds, R.~G.~Petersen, Phys.~Rev.~Lett. {\bf 2}, 499 (1959).

\bibitem{Tritt} T.M. Tritt, Thermal Conductivity: Theory, Properties, and Applications, Kluwer, New York, (2004).

\bibitem[]{Laubitz} M. J. Laubitz and T. Matsumura \textit{Canadian Jour. Phys.} \textbf{51}(12), 1247 (1973).


\end{thebibliography}
\end{document}